
\input harvmac.tex

\let\absrm\tenrm
\let\abssl\tensl

\def\to{\rightarrow}
\def\bar{\overline}
\def\MSB{{\bar{\rm MS}}}
\def\aMSB{\alpha_\MSB}

\def\CMP#1{{\sl Comm. Math. Phys.} {\bf #1}}
\def\EPL#1{{\sl Europhys. Lett.} {\bf #1}}
\def\NPB#1{{\sl Nucl. Phys.} {\bf B#1}}
\def\PLB#1{{\sl Phys. Lett.} {\bf B#1}}
\def\PRD#1{{\sl Phys. Rev.} {\bf D#1}}
\def\PRL#1{{\sl Phys. Rev. Lett.} {\bf #1}}
\def\SJNP#1{{\sl Sov. J. Nucl. Phys.} {\bf #1}}

\def\seillac{
  Proceedings of the International Symposium on Lattice Field Theory
  {\sl ``Field Theory on the Lattice''}, Seillac, France,
  {\sl Nucl. Phys.} {\bf B} {\sl (Proc. Suppl.)} {\bf 4} (1988)}.


\centerline{\bf RECENT DEVELOPMENTS IN LATTICE QCD}

\bigskip
\centerline{APOORVA PATEL\footnote{$^*$}{E-mail: adpatel@cts.iisc.ernet.in}}
\medskip
\vbox{\it\baselineskip=12pt
\centerline{CTS and SERC, Indian Institute of Science}
\centerline{Bangalore-560012, India}}

\bigskip
\centerline{ABSTRACT}
\medskip
\vbox{\absrm\baselineskip=12pt\narrower\noindent
I review the current status of several lattice QCD results.
I concentrate on new analytical developments and
on numerical results relevant to phenomenology.}

\newsec{Introduction} 

Lattice regularisation of field theories provides a framework for
their rigorously non-perturbative study from the first principle.
Such a formulation is useful for studying both fixed-point field theories
(where the lattice regulator is ultimately removed by taking the scaling
limit), and effective field theories (where the lattice regulator
acts as a fixed cutoff beyond which the theory loses its meaning).
After almost a decade in hibernation---testing algorithms, optimising
parameters, building computers---lattice studies have now reached a
stage where phenomenologically useful results are beginning to be
produced with all the systematic errors under control. I review here
some of the recent exciting developments, but also remind the reader
that there is still a long way to go and many new things to learn.

A convenient way to write down the lattice theory is in the Euclidean
path integral framework, where the integration variables are defined on
a hypercubic space-time grid and the Langrangian density is discretised
by replacing derivatives by finite differences. Then we can express
the expectation values of observables, for instance in QCD
\ref\latQCD{K. Wilson, \PRD{10} (1974) 2445
    \semi   K. Wilson, in {\it New phenomena in subnuclear physics}, Erice
            lectures 1975, ed. A. Zichichi, (Plenum, New York, 1977).},
as
\eqn\observ{
\vev{\CO} ~=~ {1 \over Z}
      \int [dU]~[d\bar{\psi}~d\psi] ~\CO~ \exp[-S_G -S_F] ~~,
}
\eqn\partfun{
Z ~=~ \int [dU]~[d\bar{\psi}~d\psi]     ~ \exp[-S_G -S_F] ~~.
}
Here the group matrices $U$ represent the gauge connection between adjacent
lattice sites, $\psi$ are the quark fields, $S_G$ and $S_F$ respectively
are the discretised gauge and fermion actions.
The simplest choice for the gauge action is the plaquette action:
\eqn\plaqact{
S_G ~=~ \beta \sum_{x,\mu<\nu} [1 - \Tr (U_{x,\mu} U_{x+\mu,\nu}
                                U_{x+\nu,\mu}^\dagger U_{x,\nu}^\dagger)]
  ~~,~~ \beta \equiv 6/g^2 ~~.
}
The two popular lattice fermion schemes are: (a) Wilson fermions
corresponding to (the hopping parameter $\kappa$ controls the quark mass)
\eqn\wilferm{
S_F^W ~=~ \sum_x \bar{\psi}_x \psi_x - \kappa\sum_{x,\mu}
          [ \bar{\psi}_x       (1-\gamma_\mu) U_{x,\mu}         \psi_{x+\mu}
          + \bar{\psi}_{x+\mu} (1+\gamma_\mu) U_{x,\mu}^\dagger \psi_x ] ~~,
}
and (b) staggered fermions corresponding to
($\chi, \eta_\mu$ are the spin-diagonalised forms of $\psi, \gamma_\mu$)
\eqn\stagferm{
S_F^S ~=~ m \sum_x \bar{\chi}_x \chi_x + \half\sum_{x,\mu}
          [ \bar{\chi}_x       \eta_\mu U_{x,\mu}         \chi_{x+\mu}
          - \bar{\chi}_{x+\mu} \eta_\mu U_{x,\mu}^\dagger \chi_x ] ~~.
}
The lattice theory does not have the same symmetry properties as the
continuum field theory. It is anticipated that the explicitly broken
symmetries (e.g. rotational and chiral) would be recovered in the
continuum limit (as the lattice spacing $a$ is taken to zero holding
the physical scale fixed), and the numerical evidence indeed points
in this direction.

The Monte Carlo importance sampling method to evaluate the path
integral is a brute force statistical analysis.
Nonetheless, it is attractive because the calculation does not have
any free parameters other than the gauge coupling and the quark masses.
The coupling $g=g(a)$ is a function of the lattice cutoff.
It is asymptotically free and determines the scaling behaviour as one
approaches the continuum limit. The quark masses are adjustable
parameters which can be freely varied between the non-relativistic
quark model case ($m\to\infty$) and the chiral limit ($m\to0$).
Such a freedom to choose parameters is a tremendous advantage in
understanding the relativistic effects and sea quark contributions.
The lattice theory does not have an inherent scale; all lattice results
come out in units of the spacing $a$. The absolute value of the lattice
cutoff $a$ has to be fixed by assigning some dimensionful physical
quantity its experimental value, and afterwords the results can be
expressed in physical units (say in GeV). In other words, only
dimensionless quantities, such as mass ratios, are uniquely predicted
in lattice calculations.

Computer simulations can of course deal with only a finite system.
The available computer power dictates the number of points that can
be simulated on a space-time grid, and then the parameters have to be
chosen so as to keep the largest correlation length well within the
finite box. To extract physical information out of such a truncated
lattice world, three crucial extrapolations are necessary:
\item{(1)}{The thermodynamic limit $L\to\infty$ removes the infrared
cutoff.}
\item{(2)}{The continuum limit $a\to0$ removes the ultraviolet cutoff.}
\item{(3)}{The chiral limit $m\to0$ allows one to reach realistic
quark masses.}

\noindent
These extrapolations help us convert results obtained on a finite
lattice, with non-zero lattice spacing and not so light quark masses
to physical numbers. They are governed by specific scaling rules---the
finite volume scaling, the renormalisation group evolution and the
chiral perturbative expansion respectively. There are two aspects to
applying these rules for removing lattice artifacts:
(i) the analytical functional forms governing the limiting behaviour,
and (ii) the maximum values of $L^{-1}$, $a$ and $m$ from where one can
safely extrapolate keeping the systematic uncertainties under control.
The details of the former have been worked out over the past several
years for many quantities of interest, while the latter have to be
determined empirically by studying lattice results obtained with
different values of $L^{-1}$, $a$ and $m$.

The numerical results have greatly improved over the years due to
innovations in designing special purpose parallel computers, in finding
optimal simulation algorithms and in estimating errors of the inherently
statistical results. The available computer power has roughly grown as
$CPU \propto e^{year}$ over the last decade. The algorithms for the pure
gauge theory have evolved to successfully face the problem of critical
slowing down. Todate, however, there is no algorithm which can simulate
the full theory of QCD (i.e. including light dynamical quarks) at a
satisfactorily fast rate using the currently available computers.
It is the combination of both analytical and numerical techniques that
has brought the subject of lattice QCD to the stage it has reached today,
and developments on both these fronts are needed to make the results
still more subtantiative in future.

\topinsert{\noindent
$$
\vbox{\hbox{\vbox{\tabskip=0pt\offinterlineskip
\def\tlr{\noalign{\hrule}}
\def\dd{{\buildrel \leftrightarrow \over D}}
\halign{\strut#&\vrule#\tabskip=0.5em&
  \hfil#\hfil&\vrule#&
  \hfil#\hfil&\vrule#&
  \hfil#\hfil&\vrule#&
  \hfil#\hfil&\vrule#\tabskip=0pt\cr\tlr
\omit&height2pt&\multispan{7}&\cr
&&\multispan{7}\hfil {\bf Present Status of Lattice QCD Calculations} \hfil&\cr
\omit&height2pt&\multispan{7}&\cr\tlr
\omit&height2pt      &&                   &&          &&            &&\cr
&& Hadronic Property && Measured Quantity && Quenched && Full Theory &\cr
\omit&height2pt      &&                   &&          &&            &&\cr\tlr
&& Light Hadron Spectrum       && Mesons and Baryons        && S  && Q  &\cr
&&                             && Glueballs                 && S  && A  &\cr
&& Topological Structure       && Susceptibility            && Q  && A  &\cr
&& Chiral Symmetry             && $\vev{\bar{q}q}$          && S  && Q  &\cr
&& Finite Temperature QCD      && $T_c$, Latent Heat,       && S  && Q  &\cr
&&                             && Screening Lengths         &&    &&    &\cr
&& Decay Constants             && $f_\pi,\ f_K,\ f_\rho$    && S  && Q  &\cr
&& $SU(3)$ Mass Splittings     &&$\vev{h|\bar{q}\lambda_8q|h}$&&S && Q  &\cr
&& Non-singlet Axial Couplings && $g_A,\ F_A,\ D_A$         && S  && Q  &\cr
&& Magnetic Moments            && Baryon Octet              && Q  && -  &\cr
&& Form Factors                && Electromagnetic           && Q  && -  &\cr
&& Twist-2 Structure Functions && $\bar{q}\dd_\mu\dd_\nu q$ && Q  && Q  &\cr
&& Singlet Scalar Coupling     && $\pi-N\ \sigma-$term      && S  && Q  &\cr
&& Singlet Axial Coupling      && $g_1$                     && A  && A  &\cr
&& $\theta_{QCD}$ Influence    && Neutron Electric          && A  && -  &\cr
&&                             && Dipole Moment             &&    &&    &\cr
&& Final State Interactions    && $I=2\quad \pi-\pi$        && S  && Q  &\cr
&&                             && Scattering                &&    &&    &\cr
&& $K^0-\bar{K^0}$ Mixing      && $B_K,\ \epsilon$          && S  && A  &\cr
&& Direct CP-violation         && $\epsilon^\prime$         && S  && A  &\cr
&& $\Delta I=\half$ Rule       && $K\to\pi\pi$ Decays       && Q  && -  &\cr
&& $D\bar{D},\ B\bar{B}$ Mixing&& $B_D,\ B_B,\ f_D,\ f_B$   && S  && -  &\cr
&& $D,\ B$ decays              && Semi-leptonic and         && Q  && -  &\cr
&&                             && Non-leptonic              &&    &&    &\cr
&& Heavy Quark Spectrum && $D,\ B,\ \psi,\ \Upsilon$ States && S  && -  &\cr
\tlr\cr}}}}
$$
\nobreak{\abssl\baselineskip=12pt\noindent Table 1:
Current status of results obtained from lattice QCD Monte Carlo simulations.
S=Stable, Q=Qualitative, A=Attempted.}
}\endinsert

Table 1 shows the current status of lattice QCD results for various
physical observables. In many cases, stable results have been obtained
within the quenched approximation. This is actually an uncontrolled
approximation, where all the vacuum polarisation quark loops are ignored
(or rather absorbed in the renormalisation of the gauge coupling).
Its computationally less intensive nature, however, has made it quite
popular. A priori one doesn't know how much to trust these results, but
in practice they often turn out to be not too far off the real numbers.
Simulations of the full theory have become feasible with the rapid
development of supercomputers, but all of the calculations so far
have been at the qualitative and exploratory level. One knows how
to attack various problems, and the formal machinery has been set up.
The future objective is to first verify and refine what we already know
about QCD from indirect methods (quark models, perturbation theory,
spectral sum rules, large$-N_c$ expansions etc.), and then proceed on
to predict unknown parameters and new phenomena.

The topics I have selected for discussion below are only a sample of the
many results obtained in lattice QCD, the selection being based on my
own judgement of the bearing the topics have on open questions and
phenomenological issues. There are also non-perturbative problems other
than QCD, where lattice technology is making important contributions
these days. These include the electro-weak sector of the standard model
(Higgs and Yukawa theories), random surfaces (quantum gravity),
correlated electron systems (high $T_c$ superconductivity) and numerous
models of statistical mechanics. The interested reader is refered to
the recent proceedings of the annual lattice field theory meetings
\ref\tallahassee{
  Proceedings of the International Symposium on Lattice Field Theory
  {\sl ``LATTICE 90''}, Tallahassee, Florida, USA,
  {\sl Nucl. Phys.} {\bf B} {\sl (Proc. Suppl.)} {\bf 20} (1991).}
\ref\tsukuba{
  Proceedings of the International Symposium on Lattice Field Theory
  {\sl ``LATTICE 91''}, Tsukuba, Japan,
  {\sl Nucl. Phys.} {\bf B} {\sl (Proc. Suppl.)} {\bf 26} (1992).}
\ref\amsterdam{
  Proceedings of the International Symposium on Lattice Field Theory
  {\sl ``LATTICE 92''}, Amsterdam, The Netherlands, to appear in
  {\sl Nucl. Phys.} {\bf B} {\sl (Proc. Suppl.)} (1993).}
for further details.

\newsec{Analytical Developments} 

\subsec{Infrared Limit of QCD Strings} 

Regge phenomenology, for instance linear trajectories in the $m^2-J$ plane
for various hadron multiplets, was the origin of string theory for hadrons.
't Hooft's $1/N$ expansion performed a topological reorganisation of
weak coupling perturbation series for $SU(N)$ gauge theories
\ref\oneoverN{G. 't Hooft, \NPB{72} (1974) 461.},
while Wilson's strong coupling expansion provided an explicit realisation
of how a lattice gauge theory may resemble a string theory \latQCD.
Forming a connection between asymptotic freedom at short distances and
an effective string picture at long distances has remained a tempting
idea for many years. Numerical results for pure gauge lattice QCD
demonstrate how the static potential between a quark-antiquark pair
changes over from the short distance Coulomb form to a long distance
linear confining form (see Fig. 3 below),
and even provide a profile of the flux tube
\ref\fluxtube{See for instance: T. Barczyk, R. Haymaker, V. Singh,
              E. Laermann and J. Wosiek, in Ref. \tsukuba, p.462.}.
The advances in conformal field theory in recent years have brought us
to a stage where we can try to put together various pieces of the puzzle
and take concrete steps towards understanding the nature of the infrared
limit of QCD strings
\ref\polchinski{J. Polchinski, talk at the Symposium on Black Holes,
                Wormholes, Membranes and Superstrings, H.A.R.C., Houston,
                hep-th 9210045 (1992).}
\ref\gliozzi{F. Gliozzi, Lectures at the XXXII Cracow School of Theoretical
             Physics, Zakopane, Poland, hep-lat 9210007 (1992).}.

A string theory of QCD has to be interpreted in an effective field theory
language without worrying about renormalisability or other ultraviolet
problems. A careful quantisation of the Nambu-Goto string away from its
critical dimension gives an effective string action, which involves the
induced (rather than Liouville) metric and which can be expanded in terms
of higher dimensional operators
\ref\effstring{J. Polchinski and A. Stronminger, \PRL{67} (1991) 1681.}.
This is not a free theory of world sheet fields $X^\mu$. In fact it is
known that the gauge theory strings have non-local contact interactions
\ref\surfinter{D. Weingarten, \PLB{90} (1980) 285.}.
Lattice strong coupling expansions show that these contact interactions
are repulsive in character, leading even to self-avoiding surfaces in
some particular cases
\ref\surfselfavoid{F. David, \EPL{9} (1989) 575
    \semi          F. David and H. Neuberger, \PLB{269} (1991) 134.}.

The high temperature behaviour of the free energy of the QCD flux tube
can be calculated
\ref\highTstring{J. Polchinski, \PRL{68} (1992) 1267.}.
A comparison of the result with that for the Nambu-Goto string shows
that the two do not agree unless more and more excitations become
available to the Nambu-Goto string with increasing temperature.
The QCD string, like the Nielsen-Olesen vortex which is also a gauge
string, is ``fat'' and the extra degrees of freedom needed may just be
the internal shape excitations. It is also possible that a more
appropriate description of the QCD string can be given in terms of
(a) a rigid string which is asymptotically free but not unitary
\ref\rigidstring{A. Polyakov, \NPB{268} (1986) 406
    \semi        H. Kleinert, \PLB{174} (1986) 335.},
(b) a string with Dirichlet boundary conditions
\ref\dirichstring{M. Green, \PLB{282} (1992) 380.}.

Gliozzi has emphasised the restrictions on the effective string
picture arising from the underlying gauge theory \gliozzi. The gauge
string carries a colour electric flux, so the effective theory must
possess a $Z_2$ automorphism corresponding to $q \leftrightarrow \bar{q}$
or $A_\mu \leftrightarrow -A_\mu$. The presence of the flux also means
that the surface is orientable and its two ends are incompatible
(a $q$ can annihilate with a $\bar{q}$ but not with another $q$).
This incompatibility implies that the lowest mode propagating along the
string world sheet cannot be the ground state with conformal weight
$h=0$; the lowest physical mode must have $h > 0$. The two ends of the
colour flux tube can be distinguished by supplementing the theory with
an additional conserved quantum number. Gliozzi uses for this purpose
the winding number of the free boson compactified on a circle with
radius $R_f$ corresponding to the finite thickness of the string. In such
a case, a free fermion mode (massive soliton) can exist in the spectrum
as an allowed topological excitation. Such a fermion obeys the gauge
theory constraints and naturally accounts for the repulsive character of
the string self-interactions. Gliozzi's conjecture is that this fermion
is the lowest excitation characterising the infrared limit of QCD string.
This situation corresponds to a $c=1$ conformal field theory where the
lowest excitation is a fermion with conformal weight $h=1/32$. Under
this assumption, dimensionless ratios such as $T_c/\sqrt{\sigma}$,
$m_G/\sqrt{\sigma}$, $R_f\sqrt{\sigma}$ have been calculated
\ref\univstring{M. Caselle, F. Gliozzi and S. Vinti, in Ref. \amsterdam.}.
They turn out to be universal numbers dependent only on the embedding
dimension of the string and independent of the gauge group, and not too
far off from the numerical lattice results. It also can be argued that
at the deconfinement temperature, the effective string theory has only
a discrete set of degrees of freedom, i.e. it becomes a topological
conformal field theory \gliozzi.

\subsec{Lattice Gauge Theory in Terms of Dual Variables} 

Sharatchandra and collaborators have carried out exact duality
transformations for lattice gauge theories. Such transformations
separate topological degrees of freedom and thus isolate the effects
of compactness of the gauge group. For example, the $U(1)$ pure gauge
theory in $3+1$ dimensions can be written as
\ref\dualuone{M. Mathur and H. Sharatchandra, {\it ``Unified Approach
              to Duality Transformations: Abelian Symmetries''},
              IMSc preprint (1992).}:
\eqn\uonepartfun{
Z ~=~ \sum_{p_\mu \in Z} \exp[ -g^2 \sum_{x,\mu\ne\nu}
      (\Delta_\mu p_{x,\nu} - \Delta_\nu p_{x,\mu})^2] ~~,
}
where $p_{x,\mu}$ is the integer valued dual vector potential living on
the links of the lattice and $\Delta_\mu$ stands for the discretised
derivative. For the more interesting case of non-Abelian gauge theories,
the Gauss's law constraint can be solved exactly in the Hamiltonian
formulation
\ref\dualsutwoH{R. Anishetty and H. Sharatchandra, \PRL{65} (1990) 813.}.
This reduces the dynamics to local gauge invariant variables which create
and annihilate a unit of colour electric flux. The dynamics then can be
recast into an Abelian gauge theory framework
\ref\abelsutwo{B. Gnanapragasam and H. Sharatchandra, \PRD{45} (1992) 1010.}.
This formulation places on a firm footing 't Hooft's conjecture
\ref\thooft{G. 't Hooft, \NPB{190} [FS3] (1981) 455.}
that topological excitations of the Abelian subgroup of $SU(N)$ determine
the confinement mechanism. The formulation is also closely related to
the discretised models of random surfaces and quantum gravity. For instance,
in the case of $SU(2)$, the lattice can be interpreted as a discretised
membrane with half-integral link lengths obeying triangle inequality of
angular momentum addition \dualsutwoH. A generalisation to the $SU(3)$
theory in terms of the integers characterising the various irreducible
representations has also been carried out
\ref\abelsuthree{R. Anishetty, G. Gadiyar, M. Mathur and H. Sharatchandra,
                 \PLB{271} (1991) 391.}.
It is also straightforward to convert the formulation to the Lagrangian
framework, e.g. the $SU(2)$ theory on a hypercubic lattice in $2+1$
dimensions can be rewritten as a sum over products of $6j-$symbols
resulting from addition of angular momenta
\ref\dualsutwoL{R. Anishetty, S. Cheluvaraja, H. Sharatchandra and M. Mathur,
                Matscience preprint IMSc-92/41, hep-lat 9211024 (1992).}.
Such explicit transformations with their elegant geometrical interpretations
need to be explored further. A combination of these techniques with
an appropriately chosen lattice (e.g. as in Ref. \surfselfavoid),
of course assuming universality, might lead to importent new results.
It would also be interesting to see how the dynamics of the $SU(2)$ gauge
theory (characterised by half-integers) differs from that of the $SO(3)$
gauge theory (characterised by integers).

\subsec{Rigorous Inequalities} 

One can derive rigorous inequalities among correlation functions
for vector-like gauge theories such as QCD.  The basis of such
inequalities is the positivity of the measure in the Euclidean
path integral:
\eqn\euclmeas{
[dA_\mu]~[d\bar{\psi}~d\psi]~\exp[-S_G -S_F] ~\ge~ 0 ~~.
}
This property has been exploited, both on the lattice
\ref\weingarten{D. Weingarten, \PRL{51} (1983) 1830.}
and in the continuum
\ref\vafawitten{C. Vafa and E. Witten, \NPB{234} (1984) 173\semi
                E. Witten, \PRL{51} (1983) 2351.},
to derive inequalities among $2-$point correlation functions.
Such inequalities yielded the result, for example, that the pion
is the lightest hadron.

These arguments can be extended to multi-point correlation functions
\ref\fourptineq{A. Patel, unpublished.}.
Consider the correlation functions of $4-$Fermion operators
between two pseudoscalar meson states at zero spatial momentum.
The Cauchy-Schwarz inequality can be applied to the ``off-shell''
correlation functions (both the mesons on the same side of the operator)
for operators with Dirac tensor structure $\Gamma\otimes\Gamma^\dagger$.
The matrix elements are then bounded from below by their values in the
Vacuum Saturation Approximation (VSA). This inequality can be extended
to the chiral limit, since the ``off-shell'' threshold amplitude is
real and the final state interactions of the psuedo-Goldstone bosons
vanish as powers of momenta in the chiral limit. Therefore, for matrix
elements which have a smooth chiral limit (e.g. the $\Delta I=3/2$ and
the $\Delta S=2$ operators fall in this class), the inequality in the
chiral limit provides a reasonable indication of the size of the
``on-shell'' matrix elements at finite quark mass.

The largest matrix elements are obviously the ones with the structure
$\gamma_5\otimes\gamma_5$. In practice, the VSA almost saturates the
matrix elements in this case
\ref\bkwil{See for instance: R. Gupta, D. Daniel, G. Kilcup, A. Patel
           and S. Sharpe, Los Alamos preprint LA-UR-91-3522,
           hep-lat 9210018 (1992).}.
Furthermore, the numerical results show that the matrix elements for
$\gamma_5\otimes\gamma_5$ are larger by an order of magnitude or more
compared to those of other Dirac tensor structures. It follows that when
$\gamma_5\otimes\gamma_5$ occurs as one of the terms amongst the various
contractions of the correlation function at tree level, it totally
overwhelms all the other terms. (Note that the matrix elements of
$\gamma_5\otimes\gamma_5$ go to a constant in the chiral limit and are
not suppressed.) The resulting $B-$parameter for the full correlation
function then is not far from $1$, even though the inequality strictly
does not hold. This is the case for the electro-penguin operators
\ref\epenguins{J. Bijnens and M. Wise, \PLB{137} (1984) 245.},
whose $B-$parameters turn out to be within $10\%$ of unity.

We also note that the correlation inequality holds for the bare
correlation functions without any reference to the cutoff scale,
while the matrix elements appearing in it may have non-vanishing
anomalous dimensions and be scale dependent. In such a case,
the anomalous dimensions must satisfy the constraint
\eqn\anomdimineq{
\gamma_{\CO \CO^\dagger} ~\ge~ 2 \gamma_{\CO} ~~,
}
so that the correlation inequality holds at any arbitrary scale.
In cases where $\CO \CO^\dagger$ and/or $\CO$ are not eigenstates
of the anomalous dimension matrix, the result of Eq.\anomdimineq\
applies to the largest anomalous dimensions occuring in the
decomposition of the operators among various eigenstates.

Similar correlation inequalities can also be applied to a system
in a finite volume \fourptineq. L\"uscher's analysis
\ref\luescher{M. L\"uscher, \CMP{104} (1986) 177; \CMP{105} (1986) 153.}
shows how the scattering length can be extracted from finite volume
correlation functions measured at two particle threshold. Explicitly,
the lowest energy of a two particle state in a sufficiently large cubic
box of length $L$ differs from twice the energy of a single particle
state by an amount proportional to the scattering length $a_0$:
\eqn\deltaE{
\delta E ~=~ E_2 - 2 E_1 ~\propto~ -{a_0 \over L^3} ~~.
}
The inequality then tells us that the interaction between two mesons in
the gluon exchange channel is attractive at threshold, i.e. the scattering
length is positive. It also can be deduced that the total threshold
interaction between two pseudoscalar mesons is attractive in the flavour
antisymmetric representation (e.g. {\bf 20} of flavor $SU(4)$).

In case of QED (which is a vector gauge theory too) the interaction
between neutral atoms is dominated by the massless photon exchange
compared to the massive electron exchange, at least at long distances.
With some additional assumptions (the atomic propagator in individual
QED field configurations is complex in general), it can be shown that
the photon exchange polarization interaction is always attractive
\fourptineq. This is an essential ingredient in understanding why
any gas condenses in to a liquid at low temperatures.

For hadrons containing a single heavy quark, the fact that the heavy
quark propagator is just a unitary matrix, can be exploited to derive
a lower bound on the $\bar{\Lambda}$ parameters of the heavy quark
effective theory
\ref\hqineq{Z. Guralnik and A. Manohar, CERN preprint CERN-TH.6766/92,
            hep-ph 9212289 (1992).}.
These $\bar{\Lambda}$ parameters are the differences between the
hadron masses and the mass parameter for the heavy quark, and they
characterise the $1/m$ corrections in the heavy quark effective theory.

\subsec{Chiral Singularity in the Quenched Approximation} 

The neglect of dynamical quark loops leaves the quenched theory non-unitary.
One of the consequences of this rather adhoc approximation is that there is
an additional unphysical pseudo-Goldstone boson in the chiral limit---the
flavour singlet $\eta^\prime$. This means that in the surrounding cloud,
the quenched hadrons have an extra $\eta^\prime$ that is absent in the
full QCD. The pseudoscalar meson cloud is an important part of the
structure of the hadrons which manifests itself at small quark masses in
terms of chiral logarithms characterising the infrared chiral singularity.
The differences in chiral logarithms between the quenched and the full
theory results can be evaluated in chiral perturbation theory by writing
the hadron propagators in terms of quark lines and omitting the diagrams
with closed quark loops
\ref\chirallog{S. Sharpe, \PRD{46} (1992) 3146; in Ref. \amsterdam.},
or by adding supersymmetric scalar ghosts to the QCD Lagrangian such that
the unphysical ghost loops cancel the effect of the closed quark loops
\ref\superchiral{C. Bernard and M. Golterman, \PRD{46} (1992) 853;
                 in Ref. \amsterdam.}.
It turns out that the unphysical behaviour of the $\eta^\prime$ gives
rise to undesirable singularities in the chiral limit of the quenched
theory. For example, the quark condensate diverges and the pseudoscalar
meson mass does not obey the GellMann-Oakes-Renner formula. The
implications of such a behaviour are not yet clear. The quenched
approximation has been of enormous use in many calculations uptodate
(mainly because at present even the best algorithms and the fastest
computers are not good enough to do a reasonable job of simulating
the full theory), so one doesn't want to discard it rightaway.
The optimistic view is that the effects of the unphysical $\eta^\prime$
cloud are negligible as long as one works above a certain quark mass.

\subsec{Lattice Artifacts, Improved Actions and Operators} 

The discretised lattice theory with a finite cutoff contains,
relative to its continuum analogue, non-universal higher dimensional
terms suppressed by powers of the lattice spacing (modulo
logarithmic renormalisations). These terms must be either removed
or made negligible by a suitable choice of lattice parameters,
before scaling the lattice results to the continuum limit.
Monte Carlo Renormalisation Group approach has shown that for
the simple plaquette action, one at least needs $\beta \ge 6.2$
to get to within $10\%$ of the asymptotic scaling behaviour
\ref\mcrgbfun{R. Gupta, G. Kilcup, A. Patel and S. Sharpe,
              \PLB{211} (1988) 132.}.
This result is also confirmed by looking at the scaling behaviour
of physical observables such as string tension, glueball masses and
the phase transition temperature. When the quarks are included,
a good criterion for judging the scaling behaviour is to look at
how fast the masses of particles belonging to the same continuum
multiplet but different lattice multiplets come together. Again
the results for the quenched theory confirm the above estimate
\ref\stagspect{S. Sharpe, R. Gupta and G. Kilcup,
               in Ref. \tsukuba, p.197.}.

In principle the lattice artifacts causing departures from
the scaling behaviour can be systematically eliminated
by improving both the action and the operators. The methodology
for a perturbative improvement program was outlined by Symanzik
\ref\pertimp{K. Symanzik, in Proceedings of {\it Trieste Workshop on
             Non-perturbative Field Theory and QCD}, Dec. 1982,
             (World Scietific, Singapore), p.61.},
while the Monte Carlo Renormalisation Group approach yields
a non-perturbatively improved action as a byproduct.
Redundant operators (i.e. the ones that can be eliminated
using the classical equations of motion) are a great convenience
in simplifying the various terms appearing in an improved action.
For the pure gauge theory, the scaling violations are $O(a^2)$,
and both perturbative
\ref\impgacta{M. L\"uscher and P. Weisz, \CMP{97} (1985) 59.}
and non-perturbative
\ref\impgactb{A. Patel and R. Gupta, \PLB{183} (1986) 193.}
approaches find that a negative admixture of $1\times2$ Wilson loop
is required to improve the scaling behaviour of the action.

Improvement for the fermions is more important,
since they have stronger scaling violations, at $O(a)$.
Moreover, the divergences arising out of explicit breaking of the
continuum symmetries on the lattice can make even the leading
scaling behaviour non-perturbative. For example, the breaking of
chiral symmetries gives rise to linear ultraviolet divergences
\ref\wilzfact{M. Bochicchio, L. Maiani, G. Martinelli, G. Rossi and
              M. Testa, \NPB{262} (1985) 331.},
which when combined with lattice artifacts formally suppressed by
powers of $a$ give corrections which are suppressed only as $O(g^2)
= O(1/\ln(a))$. These corrections are, therefore, more important
than the $O(a)$ corrections for small enough $a$. When they do not
mix the lattice operators of interest with lower dimensional
operators, they can be expressed as renormalisation $Z-$factors
relating the continuum and the lattice versions of the operators:
\eqn\defzfact{
\CO ~=~ Z_\CO^L ~ \CO^L ~~.
}
$Z_\CO^L$ are functions of the lattice spacing $a$ and should ideally
be determined non-perturbatively. They have been discussed extensively
in the literature
\ref\opzfact{See L. Maiani, G. Martinelli, M. Paciello and B. Taglienti,
             \NPB{293} (1987) 420, and references therein.},
and it turns out that they can be estimated reasonably well using the
mean field improved perturbation theory described in the following
subsection.

Next consider the $O(a)$ terms. The Wilson term in the action used for
eliminating the fermion doubling is $O(D^2 a)$ and falls in this class.
It is found that if the fermion doubling problem is removed using the
redundant interaction $\bar{\psi} (D\!\!\!/ + m)^2 \psi$ instead of the
Wilson term $\bar{\psi} D^2 \psi$, then these artifacts can be pushed
to $O(a^2)$
\ref\impwact{B. Sheikholeslami and R. Wohlert, \NPB{259} (1985) 572.}.
In addition to improving the action (or propagators), it is necessary
to have an improvement program for the operators
\ref\impwop{G. Heatlie, G. Martinelli, C. Pittori, G. Rossi and
            C. Sachrajda, \NPB{352} (1991) 266.},
to completely get rid of all the $O(a)$ terms. Such terms include
logarithmic corrections of type $O(g^2 a\ {\rm ln}a)$ arising from
renormalisations, which become $O(a)$ corrections after taking in to
account the scaling of $g^2$. For staggered fermions the scaling
violations arising from the quark propagators are already $O(a^2)$;
it is solely the lattice representation of the operators that may
give rise to $O(a)$ artifacts.

Once the lattice artifacts have been removed, the standard weak
coupling perturbation theory can be used, say at $1-$loop and at
a particular scale, to perform an appropriate matching between
the lattice and continuum renormalisation schemes. The exception
to this rule occurs when there is an unwanted mixing of lattice
operators in to lower dimensional operators. Such mixing has to be
gotten rid of non-perturbatively by imposing a physical constraint.
The Wilson fermion approach, due to lack of chiral symmetry, suffers
from this problem, e.g. the $4-$Fermion operators describing the
$\Delta I=\half$ weak decays get mixed with the quark bilinear
operators. Such a handicap is not present in the staggered fermion
approach, making it easier to relate the staggered matrix elements
to the continuum ones.

\subsec{Lattice Perturbation Theory with Tadpole Summation} 

Weak coupling perturbation theory is essential for proper
renormalisation of lattice operators so that they can be matched
with their continuum analogues. Assuming that the lattice spacing
$a$ is small, the straightforward perturbative expansion is
\eqn\uexpand{
U_\mu ~\equiv~ \exp[iagA_\mu] ~=~ 1 + iagA_\mu + \cdots ~~.
}
The quantum corrections to this expansion, however, do not vanish as
powers of $a$. On the lattice, the contractions of the $A_\mu$'s with
each other generate ultraviolet divergences which can cancel the
additional powers of $a$. The most troublesome of these contractions
are the quadratically divergent tadpole diagrams, which are absent in
the continuum but present on the lattice. The quadratic divergence
precisely cancels the powers of $a$ accompanying $g$, leaving behind
terms which are suppressed only by powers of $g^2$ and not of $a$.
Lepage and Mackenzie have proposed a mean field method to suppress
these uncomfortably large corrections
\ref\tadpolesum{G. Lepage and P. Mackenzie, in Ref. \tallahassee, p.173;
                Fermilab preprint FERMILAB-PUB-19/355-T,
                hep-lat 9209022 (1992).}.
In this method the tadpoles are summed up modifying Eq.\uexpand\ to
\eqn\uexpandmf{
U_\mu ~=~ u_0(1 + iagA_\mu + \cdots) ~~,
}
with the convention that all tadpole diagrams are to be dropped from the
perturbative calculations of renormalisation constants. $u_0$ here is
a non-perturbative parameter to be taken from the numerical simulations;
convenient choices are the Landau gauge expectation value of the gauge
link, or the fourth root of the expectation value of the plaquette.
Rewriting the lattice gauge action so as to make the factors of $u_0$
explicit, one finds that the effective gauge coupling to be used in
perturbative calculations becomes $g^2_{eff} = g^2/u_0^4$. Similarly
the continuum fermion fields are better described by the operators
$\sqrt{2u_0 \kappa} \psi$, $\sqrt{u_0} \chi$.

This rewriting of perturbation theory has brought the numerical data
for $g^2 \approx 1$ in good agreement with scaling \tadpolesum;
asymptotic scaling in terms of the bare lattice coupling fails
miserably for these data.

\subsec{Renormalizations in Heavy Quark Effective Theory} 

In the limit of infinite quark mass, the dynamics of QCD is invariant
under spin-flavour symmetries of the heavy quark degrees of freedom.
This has been exploited by Isgur and Wise, for hadrons containing a
single heavy quark, to relate many form factors and to reduce many
matrix elements down to a few unknown functions
\ref\isgurwise{N. Isgur and M. Wise, \PLB{232} (1989) 113;
               \PLB{237} (1990) 527.}.
Despite the impressive simplifications arising in the limiting case,
it is a must to estimate the size of the $O(1/m)$ corrections to the
leading terms, in order to attach proper meaning to the results and
obtain reliable predictions for physical heavy quark states. The
virtue of the formalism of heavy quark effective theory, developed
over the past few years, lies in providing a model independent
$m\to\infty$ limit, which can be systematically improved with
power series expansions in $1/m$
\ref\hqefftheory{An overview of the entire subject can be found in:
                 {\it Heavy Flavours}, ed. A. Buras and M. Lindner,
                 Advanced Series on Directions in High Energy Physics,
                 (World Scientific, Singapore, 1992).}.

In the continuum, the problem is commonly studied using the static
(i.e. constant velocity frame) formulation. This language can handle
the $O(1/m)$ corrections but not the $O(\alpha_s/m)$ ones, and hence
cannot be applied to hadrons containing more than one heavy quark
\ref\heavyqrev{See for instance: E. Eichten, in Ref. \tallahassee, p.475.}.
On the lattice, however, there is no such restriction on the dynamics
of the heavy quark. The simplest choice is to just use the standard
fermion action with a large mass and study the mass dependence of
various matrix elements. In such a case, as explained below, proper
normalisation factors have to be included to obtain sensible results.

At finite lattice spacing, even for a free field theory, the mass
parameter in the Lagrangian does not agree with the position of the
pole in the propagator. This mismatch has to be eliminated when the
matrix elements are extracted using the LSZ reduction formula.
In the mean field theory approach, a tree level wavefunction
renormalisation is sufficient to get rid of the problem
\ref\heavywfren{M. L\"uscher, \CMP{54} (1977) 283
    \semi       G. Lepage, in Ref. \tsukuba, p.45
    \semi       A. Kronfeld, in Ref. \amsterdam.}.
Combining this correction factor with the tadpole summation factor
discussed in the previous subsection, the effective fermion masses
to be matched with a continuum theory become
\eqn\effmq{
m_{eff}^W ~=~ \ln[1 + {1 \over 2u_0}({1\over\kappa} - {1\over\kappa_c})]
{}~~,~~ m_{eff}^S ~=~ \sinh^{-1}(m/u_0) ~~,
}
while the effective lattice fields representing the continuum fermions
take the form
\eqn\effwf{
\psi_{eff} ~=~ [1 + 2u_0\kappa - (\kappa/\kappa_c)]^{1/2} \psi ~~,~~
\chi_{eff} ~=~ [u_0^2 + m^2]^{1/4} \chi ~~.
}
Clearly the effect of the correction factors becomes more and
more important as the lattice masses increase. It is easy to see,
for example using a hopping parameter expansion, that one must
incorporate these factors to obtain the correct values for the
matrix elements in the infinite quark mass limit
\ref\heavyqmeren{S. G\"usken, K. Schilling, R. Sommer, K.-H. M\"utter
                 and A. Patel, \PLB{212} (1988) 216.}.

An alternative is to use the non-relativistic expansion of the
fermion action with the expansion coefficients determined by
appropriate matching conditions
\ref\nrqcdlat{G. Lepage and B. Thacker, in \seillac, p.199.
    \semi     B. Thacker and G. Lepage, \PRD{43} (1991) 196.}.
The mass term is dropped altogether, and the matrix elements
of leading as well as subleading operators are measured in the
effective theory. This effective theory, however, is not
renormalisable and contains power law divergences. This feature
is reflected in non-perturbative contributions to the expansion
coefficients. The debate on how well these coefficients can be
estimated is not settled as yet.

The mean field prescription is able to get rid of lattice artifacts
which are $O(ma)$, but not those which are $O(\Lambda_{QCD} a)$.
Thus fractional errors of $O(\Lambda_{QCD}/m)$ are left behind.
Maiani {\it et al.} argue that the only way out is to fix the
coefficients using non-perturbative matching conditions
\ref\nonpertmatch{L. Maiani, G. Martinelli and C. Sachrajda,
                  \NPB{368} (1992) 281.},
in general reducing the predictive power of the theory.
Lepage {\it et al.} argue that the power law divergences do not
give rise to large non-perturbative corrections for the couplings
typically used in lattice simulations---by restricting $m$ to be
of order $1/a$ or smaller and using the mean field improved
perturbation theory to estimate the coefficients, the fractional
error can be reduced to the level of a few percent
\ref\imppertmatch{G. Lepage, in Ref. \tsukuba, p.45.}.
The issue should get resolved when the tests of the mean field
improved perturbation theory \tadpolesum\ are pushed to the
stage where non-perturbative terms show up.

There is yet another approach to the problem possible, based on the
fact that the heavy quark effective theory treats space and time
asymmetrically. Perhaps a better description of the continnum physics
can be obtained if space and time hopping are treated differently
on the lattice too
\ref\hqetlat{A. Kronfeld, in Ref. \amsterdam
    \semi    A. Kronfeld and P. Mackenzie, in preparation.}.
This is a topic requiring further study.

\newsec{Spectral Results} 

In Euclidean (imaginary) time, particle propagators evolve as
$\exp(-E\tau)$, so the farther they propagate the more they get
dominated by the lowest energy eigenstates. It is therefore
straightforward to extract the masses of the lowest eigenstates
from the asymptotic behaviour of propagators. A few excited states
can also be handled, as long as they are stable against strong decay,
by applying variational techniques to the wavefunctions representing
the creation/annihilation operators. Unstable states (i.e. resonances)
are not easy to deal with, and no satisfactory solution to extraction
of their properties exists yet. Some progress has been made though in
relating the behaviour of multi-particle states in a finite volume to
properties of resonances
\ref\euclreson{M. L\"uscher, \NPB{364} (1991) 237.}.

Considerable improvement in the signal to noise ratio can be achieved,
if the operators are designed so as to maximise their overlap with
the desired states and reduce the contamination from excited states.
It is here where one's intuition about hadronic wavefunctions (based
on phenomenological models) is helpful, and a lot of progress has been
made in this direction in recent years. Wavefunctions can be directly
determined on the lattice using spatially extended hadron operators.
For heavy-light mesons, Coulomb gauge wavefunctions evaluated for a
spinless relativistic quark moving in the static $q\bar{q}$ potential
(which can be calculated on the same lattices) are excellent
approximations to the actual distributions
\ref\heavywf{A. Duncan, E. Eichten and H. Thacker, in Ref. \amsterdam.}.
For light hadrons, Coulomb gauge wavefunctions provide a reasonable
description of their internal structure
\ref\lightwf{M. Hecht and T. DeGrand, \PRD{46} (1992) 2155,
             and references therein.},
but there is enough room for finding a better prescription. It should be
noted that even though such wavefunctions correspond to an uncontrolled
truncation in a non-abelian gauge theory, they are a useful starting
point for understanding the hadron structure and they certainly improve
the signal to noise ratio.

A typical set up in computing matrix elements is to first calculate the
Green's functions with the desired interaction operator and the appropriate
incoming and outgoing states, let the time separations become large enough
so that the lightest states saturate the correlation functions, and then
remove the factors corresponding to the external legs following the LSZ
reduction technique. Schematically,
\eqn\generalme{\eqalign{
\vev{\CO_f (\tau_f) | \CO_{int} | \CO_i (\tau_i)} ~&=~ \sum_{j,k}
  \vev{\phi_j | \CO_{int} | \phi_k} ~ e^{-E_j \tau_f} e^{E_k \tau_i}
  ~ \vev{\CO_f | \phi_j} ~ \vev{\phi_k | \CO_i} \cr
&{\scriptstyle \tau_f \to \infty \atop {\displaystyle \longrightarrow
  \atop \scriptstyle \tau_i \to -\infty}}
  \vev{h_f | \CO_{int} | h_i} ~ e^{-E_f \tau_f} e^{E_i \tau_i}
  ~ \vev{\CO_f | h_f} ~ \vev{h_i | \CO_i} ~~. \cr
}}
Here $\phi_j(\phi_k)$ denote all the states consistent with the
quantum numbers corresponding to the annihilation (creation) operators
$\CO_f(\CO_i)$, and $h_f(h_i)$ are the lowest ones amongst them.

\subsec{Light Hadron Spectrum} 

The most easily calculable hadron masses on the lattice are those
of the pseudoscalar and vector mesons and the spin-$\half$ baryon.
It has become customary to denote them as $m_\pi$, $m_\rho$ and
$m_N$ ($N$ stands for the nucleon) repsectively. These masses are
determined as function of the quark mass on the lattice, and a
convenient way to represent the results is in terms of dimensionless
mass ratios. One such plot, called the APE plot, depicting the
behaviour of $m_N/m_\rho$ vs. $(m_\pi/m_\rho)^2$ is shown in Fig. 1.
With increasing quark mass, $m_\pi/m_\rho$ monotonically varies from
$0$ to $1$, each particular value for $m_\pi/m_\rho$ representing
a theory for which all other dimensionless ratios can be determined.

\topinsert{
\vskip 15truecm
\medskip
{\abssl\baselineskip=12pt\noindent Figure 1:
APE mass ratio plot for the quenched Wilson fermion spectrum.}
}\endinsert

The data shown in Fig. 1 are for quenched Wilson fermions
\ref\gfelevenspect{F. Butler, H. Chen, A. Vaccarino, J. Sexton and
                   D. Weingarten, in Ref. \amsterdam.}
\ref\apespectA{S. Cabasino {\it et al.} (APE collaboration),
               \PLB{258} (1991) 195.}
\ref\apespectB{M. Guagnelli {\it et al.} (APE collaboration),
               \NPB{378} (1992) 616.}.
The lattice parameters for these simulations are shown in Table 1,
together with the parameters of other recent calculations. They
indicate, in addition to the size of the state of the art lattice
simulations, our present understanding of the cutoff limits
$a_{max}$ and $L_{min}$ for reliable extrapolations to the continuum.

The solid curves in the figure, constrained so as to pass through
the physical points, represent the qualitative expectations based on
phenomenological models. The curve on lower left is the behaviour
expected from chiral perturbation theory, while the curve on the
upper right is the trend in the heavy quark mass expansion.
The important questions are how close the lattice results are to the
phenomenological expectations and whether the lattice numbers can be
extrapolated to the experimental point, indicated by a ``?'' in the
figure. Before looking at various systematic differences, it should
be noted that the quenched theory results do not have to agree with
the real world---a priori we do not know what the results should
look like. For example, the pion cloud around the hadrons is quite
different in the quenched approximation compared to the real world.
Also the quenched rho cannot decay and its mass can differ from the
real rho mass by an amount comparable to its width
\ref\rhodecay{P. Geiger and N. Isgur, \PRD{41} (1990) 1595.}.
Thus the apparent agreement, within $\approx 10\%$, between the queched
lattice results and phenomenological expectations has to be taken with
caution. On one hand, it is a good sign that quenched lattice QCD is
not completely off the mark. On the other hand, it may be just a lucky
coincidence for these particular variables, which would not hold for
some other observables (cf. subsection 2.4 and section 5).

There are several technical issues involved in understanding the lattice
data. Contamination from excited states is present in the correlators
and has to be carefully eliminated in order to extract the asymptotic
mass values. The nucleon, with a small gap between the lowest and
the first excited state, is particularly susceptible to this problem.
Improved operators help, but even then the truly asymptotic signal
may not be easy to get to
\ref\qcdpaxspect{Y. Iwasaki, K. Kanaya, S. Sakai, T. Yoshi\'e, T. Hoshino,
                 T. Shirakawa and Y. Oyanagi, in Ref. \amsterdam.}
\ref\lanlspectW{D. Daniel, R. Gupta, G. Kilcup, A. Patel and S. Sharpe,
               \PRD{46} (1992) 3130.}.
This effect has led to a variation amongst the numerical results quoted
by various groups. The situtation is expected to get clarified soon
with the use of sophisticated hadron wavefunctions which couple more
strongly to the ground states.

\topinsert{\noindent
$$
\vbox{\hbox{\vbox{\tabskip=0pt\offinterlineskip
\def\tlr{\noalign{\hrule}}
\halign{\strut#&\vrule#\tabskip=0.5em&
  \hfil#\hfil&\vrule#&
  \hfil#\hfil&\vrule#&
  \hfil#\hfil&\vrule#&
  \hfil#\hfil&\vrule#&
  \hfil#\hfil&\vrule#&
  \hfil#\hfil&\vrule#\tabskip=0pt\cr\tlr
\omit&height2pt&&       &&      &&            &&              &&
&&\cr
&&  Ref.       &&$\beta$&& Size && $a$(Fermi) && $\pi/a$(GeV) && $L$(Fermi)
&\cr
\omit&height2pt&&       &&      &&            &&              &&
&&\cr\tlr
&& \gfelevenspect && 5.93 &&$24^3\times36$        && 1/9  && 5.7 && 2.7 &\cr
\omit&height2pt   &&      &&                      &&      &&     &&    &&\cr
&& \apespectA     && 6.00 &&$24^3\times32$        && 1/10 && 6.3 && 2.4 &\cr
\omit&height2pt   &&      &&                      &&      &&     &&    &&\cr
&& \qcdpaxspect   && 6.00 &&$24^3\times54$        && 1/10 && 6.3 && 2.4 &\cr
\omit&height2pt   &&      &&                      &&      &&     &&    &&\cr
&& \gfelevenspect && 6.17 &&$30\times32^2\times40$&& 1/12 && 7.5 && 2.6 &\cr
\omit&height2pt   &&      &&                      &&      &&     &&    &&\cr
&& \apespectB     && 6.30 &&$24^3\times32$        && 1/16 && 10. && 1.5
&\cr\tlr
\omit&height2pt   &&      &&                      &&      &&     &&    &&\cr
&& \ref\apespectS{S. Cabasino {\it et al.} (APE collaboration),
                  \PLB{258} (1991) 202.}
                  && 6.00 &&$24^3\times32$        && 1/10 && 6.3 && 2.4 &\cr
\omit&height2pt   &&      &&                      &&      &&     &&    &&\cr
&& \ref\lanlspectS{R. Gupta, G. Guralnik, G. Kilcup and S. Sharpe,
                  \PRD{43} (1991) 2003.}
                  && 6.00 &&$24^3\times40$        && 1/10 && 6.3 && 2.4 &\cr
\omit&height2pt   &&      &&                      &&      &&     &&    &&\cr
&& \stagspect     && 6.20 &&$32^3\times48$        && 1/13 && 8.1 && 2.5 &\cr
\omit&height2pt   &&      &&                      &&      &&     &&    &&\cr
&& \stagspect     && 6.40 &&$32^3\times48$        && 2/35 && 11. && 1.8
&\cr\tlr
\cr}}}}$$
\nobreak{\abssl\baselineskip=12pt\noindent Table 2:
Various lattice parameters of recent quenched hadron spectrum calculations.
The upper and lower halves of the table correspond to Wilson and staggered
fermion calculations respectively.}
}\endinsert

Due to lattice artifacts, we typically have
\eqn\massratio{
(m_N/m_\rho)_{\rm latt} ~=~ (m_N/m_\rho)_{\rm cont} [1 + O(a)] ~~.
}
The lattice data should thus converge towards a universal curve as
$a \to 0$. Also, the masses are shifted from their infinite volume
values, if the lattice size $L$ is too small. The nucleon is
physically larger than the mesons and hence more easily distorted
by the finite box size. Empirical evidence over the last few years
has shown that both the finite lattice spacing and finite volume
effects shift the APE curve upwards.

Fig. 1 shows that for the usual Wilson fermions, with $\beta \ge 6.0$,
the scaling violations in mass ratios are reasonably under control.
More encouraging results have been obtained with an improved Wilson
fermion action. Ref.
\ref\apespectI{M.-P. Lombardo, G. Parisi and A. Vladikas,
               Rome preprint ROM2F-92-46, hep-lat 9206023 (1992).}
finds that with an $O(a)$ improved action, the scaling violations
in mass ratios are substantially reduced between $\beta=5.7$ and
$6.0$. There is not much to be gained with the improved action,
however, for $\beta \ge 6.2$
\ref\ukqcdspectI{C. Allton {\it et al.}, \PLB{284} (1992) 377.}.
This behaviour is in agreement with the logic that by using a more
complicated action one can approach the continuum limit faster.
The prospects for staggered fermions are not so bright; the scaling
violations, though formally only of $O(a^2)$, are quite large at
$\beta=5.7$ \stagspect.

An example of the finite volume effect on the hadron masses
calculated with dynamical quarks is shown in Fig. 2
\ref\kekspect{M. Fukugita, H. Mino, M. Okawa and A. Ukawa,
              \PRL{68} (1992) 761
    \semi     M. Fukugita, H. Mino, M. Okawa, G. Parisi and A. Ukawa,
              \PLB{294} (1992) 380.}.
Asymptotically, the approach to the thermodynamic limit is dictated
by the behaviour of the lightest states in the system---the pions.
The corrections can be estimated as pion exchanges between nearest
neighbour periodic images and behave as $\exp(-m_\pi L)$
\ref\finvolmass{M. L\"uscher, \CMP{104} (1986) 177
    \semi       J. Gasser and H. Leutwyler, \PLB{184} (1987) 83;
                \PLB{188} (1987) 477.}.
In the intermediate range, however, the hadron wavefunctions are
substantially squeezed. The corrections then are dominated by quark
exchanges between all the images, and depend on the sign of the quark
boundary conditions. The point particle description has to be softened
by form factors and the dominant contribution comes from the zero mode,
making the corrections behave like $1/L^3$ \kekspect.

\topinsert{
\vskip 18truecm
\medskip
{\abssl\baselineskip=12pt\noindent Figure 2:
Finite volume dependence of hadron masses for two flavours of
dynamical staggered quarks at $\beta=5.7$ \kekspect. The thin
lines are the predictions of the asymptotic virtual pion exchange
formula, while the thick lines are fits using $1/L^3$ corrections
arising from wavefunction squeezing.}
}\endinsert

Putting all the evidence together, the numerical results show that
in the quenched theory one can safely extrapolate to the continuum
for $\beta \ge 6.2$. In the absence of detailed results, it is not
yet possible to give such a precise bound for the full theory.
The hadron masses get to within one or two percent of their infinite
volume limits, for quarks heavier than half the strange quark mass,
once the lattice size is more than $2.5$ Fermi across \kekspect
\ref\milcspect{C. Bernard, T. DeGrand, C. DeTar, S. Gottlieb, A. Krasnitz,
               R. Sugar and D. Toussaint, in Ref. \amsterdam.}.
For lighter quarks, barring the possibility of resonances, extrapolations
made using chiral perturbation theory should be reliable. Having brought
all the systematic errors within a few percent, it is merely a question of
beating down the statistical errors by running a powerful enough computer
for a long enough period and obtain quantitative predictions.

\subsec{QCD Running Coupling $\alpha(q)$} 

In principle, it is straightforward to extract the QCD running coupling
from lattice calculations. One picks a value for the bare lattice
coupling $\beta$, calculates on the lattice a physically measurable
dimensionful quantity (e.g. $f_\pi$), and extracts the value for the
lattice cutoff $a$ by comparing the lattice and the physical numbers.
This procedure yields the running coupling $g^2(a)$ which can be evolved
to other scales as well as converted to other regularisation schemes
using perturbation theory.

This straightforward procedure has two important caveats requiring
careful treatment:
\item{(1)}{Due to lattice artifacts the relation between lattice and
continuum results has unphysical and non-universal corrections, e.g.
\eqn\fpilatcont{
(f_\pi)_{\rm latt} ~=~ (f_\pi)_{\rm cont} a (1 + O(a)) ~~.
}
These scaling violations preclude a naive matching of lattice and
continuum numbers.}
\item{(2)}{The perturbation theory in the bare lattice coupling is not
reliable. The problem is exemplified by the following relation between
the couplings in the continuum $\MSB$ and the lattices schemes
\ref\hasenfratz{A. Hasenfratz and P. Hasenfratz, \PLB{93} (1980) 165
    \semi       R. Dashen and D. Gross, \PRD{23} (1981) 2340.}
\eqn\alphamsbarlat{
{1 \over \aMSB(\pi/a)} ~=~ {1 \over \alpha_L(a)} - 3.880 + O(\alpha) ~~.
}
Since $\alpha_L^{-1} \approx 4\pi$, the first order correction is
rather large and higher order terms cannot be just ignored.}

The first problem can be alleviated to some extent by improving the
lattice action and operators as discussed in subsection 2.5 above.
Still the remaining unwanted lattice artifacts (terms suppressed
by various powers of $a$ and $g^2$) have to be removed by
performing simulations at a variety of bare lattice couplings,
confirming that the scaling violations are of the expected nature,
and then extrapolating the results to the continuum limit $a \to 0$.

The second problem makes it mandatory that all the results be expressed
in an appropriate renormalisation scheme, where the higher order terms
are better behaved. (This is similar in spirit to the shift from the
MS to the $\MSB$ scheme in continuum calculations, and despite the
fact that perturbation theory only provides an asymptotic series.)
As discussed in subsection 2.6, the large coefficients originate mainly
from the gauge field tadpole diagrams. Lepage and Mackenzie have argued
\tadpolesum\ that a continuum-like coupling constant, e.g. $\aMSB$ or
$\alpha_{\rm MOM}$, with the scale $q \approx \pi/a$ would be a good
expansion parameter. Their prescription is to cast all the perturbative
expressions in terms of such a coupling constant, which considerably
reduces the higher order coefficients, and then determine the coupling
constant through a non-perturbatively measured lattice quantity. One
simple choice is \tadpolesum:
\eqn\alphamsbarplaq{
{1 \over \aMSB(\pi/a)} ~=~ {1 \over \alpha_P(a)} - 0.51 + O(\alpha)
{}~~,~~ \alpha_P ~=~ -{{3 \ln \langle \Tr U_P \rangle} \over 4\pi} ~~.
}
where $U_P$ is the product of gauge links around an elementary plaquette.

These steps have been followed in Ref.
\ref\wupalpha{G. Bali and K. Schilling, in Ref. \amsterdam.}
\ref\ukqcdalpha{S. Booth {\it et al.} (UKQCD collaboration),
                \PLB{294} (1992) 385.}
to extract the running coupling of pure gauge QCD, $\aMSB^{(0)}$,
from the static $q\bar{q}$ potential, $V(R)$. Fig. 3 illustrates the
accuracy to which $V(R)$ has been determined in lattice simulations,
and its remarakable agreement with a simple power series expansion in
the $q\bar{q}$ separation $R$. The running coupling $\alpha_V (R^{-1})$
can be extracted from the short distance Coulomb behaviour of the
potential, $V(R) \propto -\alpha_V (R^{-1})/R$. The running coupling
$\alpha(\pi/a)$ can also be independently evaluated from the scaling
of the string tension (extracted from the long distance part of the
potential calculated at a variety of bare lattice couplings $\beta$).
Both these determinations coincide nicely \wupalpha. Indeed the fact
that a single lattice simulation covers both perturbative and
non-perturbative aspects of QCD, is an excellent demonstration that
the same theory can successfully explain both the high energy jet
physics as well as the low energy hadronic properties.

\topinsert{
\vskip 11truecm
\medskip
{\abssl\baselineskip=12pt\noindent Figure 3:
Static $q\bar{q}$ potential as a function of their separation
\wupalpha. The lattice coupling is $\beta=6.4$, and the short distance
lattice artifacts violating rotational symmetry have been removed by
subtracting from the lattice results the difference between lattice
and continuum $1-$gluon exchange potentials.}
}\endinsert

The outstanding drawback of these results for the pure gauge QCD is
that there is no direct way to relate the zero quark flavour coupling
constant, $\aMSB^{(0)}$, to the real coupling constant involving a
number of light quark flavours, $\aMSB$. Such a relation necessarily
involves non-perturbative physics, and at best one can estimate it
using phenomenological models. (For example, assigning the pure gauge
theory string tension a phenomenological value, $\sqrt{\sigma} = 0.44$
GeV, is only an educated guess.) The Fermilab group have made such an
estimate for the coupling determined using the spin-averaged $1P-1S$
splitting in the Charmonium system
\ref\fnalalpha{A. El-Khadra, G. Hockney, A. Kronfeld and P. Mackenzie,
               \PRL{69} (1992) 729.}.
Charmonium is described well by potential models, so in essence one
has to match the potentials of the full and the quenched theory at a
distance corresponding to the Charmonium size, $R_{c\bar{c}} \sim 0.5$
Fermi. The two potentials then differ at shorter distances, and the
difference can be estimated using the renormalisation group evolution.
Since the quenched $\beta-$function is a bit too large, the short
distance quenched coupling is a bit too small, i.e. the quenched
potential is a bit too shallow. This calculation also has some technical
advantages: (a) the orbital splitting is known to be quite insensitive
to the value of the quark mass, so no careful tuning of the quark mass
is necessary; (b) the quarkonium system is physically smaller than the
light hadrons, making it easier to control finite size effects and to
bound uncertainties in perturbative renormalisation group evolution.

All the results, converted to the $\MSB$ scheme and evolved to the
energy scale $M_Z$, are presented in Table 3 for comparison with the
experimental numbers
\ref\pdg{K. Hikasa {\it et al.} (Particle Data Group), \PRD{45} (1992) S1.}
\ref\alphalep{S. Banerjee, these proceedings.}.
It is seen that the lattice results lie slightly below the experimental
ones. Not much should be made of the small discrepancy, which is likely
to disappear as the lattice simulations improve and the dominant
systematic uncertainty in the conversion from $\aMSB^{(0)}$ to $\aMSB$
reduces. Rather the close agreement between the experimental and lattice
results, both in the value and in the size of the error, is a triumph
of lattice QCD calculations.

\topinsert{\noindent
$$
\vbox{\hbox{\vbox{\tabskip=0pt\offinterlineskip
\def\tlr{\noalign{\hrule}}
\halign{\strut#&\vrule#\tabskip=0.5em&
  \hfil#\hfil&\vrule#&
  \hfil#\hfil&\vrule#&
  \hfil#\hfil&\vrule#&
  \hfil#\hfil&\vrule#\tabskip=0pt\cr\tlr
\omit&height2pt&&            &&                    &&             &&\cr
&&  Ref.       && Observable && $\aMSB^{(0)}(M_Z)$ && $\aMSB(M_Z)$ &\cr
\omit&height2pt&&            &&                    &&             &&\cr\tlr
&& \pdg        && Experiment           &&    ---    && 0.1134(35)  &\cr
\omit&height2pt&&                      &&           &&            &&\cr
&& \alphalep   && Experiment           &&    ---    && 0.123(4)    &\cr
\omit&height2pt&&                      &&           &&            &&\cr
&& \fnalalpha  && $M_{1P}-M_{1S}$      && 0.0790(9) && 0.105(4)    &\cr
\omit&height2pt&&                      &&           &&            &&\cr
&& \wupalpha   && $q\bar{q}$ Potential && 0.0796(3) &&    ---      &\cr
\omit&height2pt&&                      &&           &&            &&\cr
&& \ukqcdalpha && $q\bar{q}$ Potential && 0.0801(9) &&    ---      &\cr
\omit&height2pt&&                      &&           &&            &&\cr\tlr
\cr}}}}$$
\nobreak{\abssl\baselineskip=12pt\noindent Table 3:
Comparison of lattice results for $\aMSB$ with experiment.
Errors shown are statistical.}
}\endinsert

In the calculations described above, a single simulation had to
simultaneously obey the requirements for $a_{max}$ and $L_{min}$.
This limits the energy range over which the running coupling can
be studied, since the lattice dimensions are constrained by the
available computer power. L\"uscher and collaborators have proposed
an approach based on discrete renormalisation group to extend this
energy range
\ref\sutwoalpha{M. L\"uscher, R. Narayanan, P. Weisz and U. Wolff,
                \NPB{384} (1992) 168
    \semi       M. L\"uscher, R. Sommer, U. Wolff and P. Weisz,
                DESY preprint DESY 92-096, hep-lat 9207010 (1992).}.
In this approach, computation of a non-perturbatively defined coupling
constant on lattices of size $L$ and $2L$, for the same bare coupling,
provides an integral of the $\beta-$function between scales $L$ and $2L$.
Short distance lattice artifacts are eliminated by applying finite
size scaling techniques to several pairs of results with sizes $L/a$
and $2L/a$ and then extrapolating to $a \to 0$. By using multiple
steps of this type, the separation between the low energy end (where
a non-perturbative input such as the string tension sets the overall
scale) and the high energy end (where perturbative scaling can be
applied with precision) is extended. In the intervening range, the
running coupling is determined non-perturbatively. Encouraging results
have been obtained in this manner for the pure gauge $SU(2)$ theory
\sutwoalpha.

\subsec{$I=2$ Pion Scattering Length} 

An understanding of the behaviour of multi-particle states below
inelastic threshold can be obtained by turning the limitation of
finite lattice volume in to an advantage. When the volume of the
system is large enough so that the particle wavefunctions are not
badly distorted by the finite box size, the shifts in the energy
levels of two-particle states from their infinite volume limit
are related to their scattering phase shifts
\ref\scatlen{M. L\"uscher, \CMP{105} (1986) 153; \NPB{354} (1991) 531.}.
In the case of two pions in a finite box, the interaction energy at
threshold is
\eqn\dEtwopi{
\delta E ~=~ E_{2\pi} - 2m_\pi ~=~ - {4\pi a_0^I \over m_\pi L^3}
                                     \bigl( 1+O({1\over L}) \bigr) ~~,
}
where $a_0^I$ is the $s-$wave scattering length for the two pions in
an isospin $I$ state. Numerically, the interaction energy is small
in the region where this formula applies, and an easy way to extract
$\delta E$ is from the ratio
\eqn\dEratio{
{G_{2\pi} (0;\tau) \over (G_\pi (0;\tau))^2}
                  ~\propto~ e^{-\delta E~\tau} ~~.
}
At present, the lattice data only allow extraction of the $I=2$
scattering length, in which case quark annihilation diagrams are
absent. (The annihilation diagrams, which are a necessary ingredient
for the $I=0$ scattering interaction, are in general noisy. Moreover,
they are likely to suffer from large systematic uncertainty in the
quenched approximation.) By extending the flavour symmetry group to
$SU(4)$ on the lattice, scattering lengths in all possible flavour
representations can be calculated. The calculations have been carried
out for both staggered and Wilson fermions
\ref\itwostag{S. Sharpe, R. Gupta and G. Kilcup, \NPB{383} (1992) 309.}
\ref\itwowil{R. Gupta, D. Daniel, G. Kilcup, A. Patel and S. Sharpe,
             Los Alamos preprint LA-UR-92-3641, hep-lat 9301016 (1993).},
and the results compare well with the predictions of lowest order
chiral perturbation theory
\ref\weinberg{S. Weinberg, \PRL{17} (1966) 616.},
e.g.
\eqn\Itwocpt{
4\pi a_0^{I=2} f_\pi^2 /m_\pi ~=~ - 1/4 ~~.
}
In fact, somewhat surprisingly, the deviations from the chiral
predictions are found to be rather small even for pseudoscalar
meson masses of $\sim 700$ MeV. Comparison of staggered and
Wilson fermion results also provides a test of how well the
current algebra is restored for Wilson fermions as $a \to 0$.

\subsec{Heavy-Light Pseudoscalar Decay Constant} 

The decay constant $f_P$ for a heavy-light pseudoscalar meson of mass
$m_P$ happens to be a good testing ground for the ideas of heavy quark
effective field theory. In case of QCD with $N_f$ light quark flavours,
it can be expanded as
\ref\fbscaling{M. Voloshin and M. Shifman, \SJNP{45} (1987) 292.}
\eqn\fbexpansion{
\phi_P ~\equiv~ f_P \sqrt{m_P} \left[ {\alpha(m_P) \over \alpha(m_B)}
                          \right]^{6/(33-2N_f)}
       ~=~ \phi_\infty \left( 1 + {A \over m_P} + {B \over m_p^2} + \cdots
                       \right) ~~.
}
Here $\phi_\infty$ is the limiting value and $A$, $B$ are constants
except for a weak logarithmic dependence on $m_P$. The question of
phenomenological importance is the size of the $1/m_P$ corrections
at the physical $B$ and $D$ masses.

Lattice QCD calculations have mapped out the behaviour of $\phi_P$
vs. $1/m_P$ as illustrated in Fig. 4
\ref\uclafb{C. Bernard, J. Labrenz and A. Soni, in Ref. \amsterdam.}.
As a function of $1/m_P$, $\phi_P$ shows a sizeable and negative
departure from its asymptotic value---there is about a factor of
two variation between the $m=\infty$ limit and the $D$ mass.
This feature is consistent with observations in the QCD spectral
sum rule approach
\ref\fpssumrul{T. Aliev and V. Eletski\u i, \SJNP{38} (1983) 936.}.

\topinsert{
\vskip 15truecm
\medskip
{\abssl\baselineskip=12pt\noindent Figure 4:
The decay constant for a heavy-light pseudoscalar meson ($\bar{Q}q$),
rescaled to take in to account the leading mass dependence and the
anomalous dimension, as a function of the pseudoscalar mass inverse
\uclafb. The $M_P^{-1} = 0$ value is from the static approximation.
The uncorrected points refer to the results prior to including the
normalisation constants of subsection 2.7.}
}\endinsert

As far as the results for $f_B({\rm static}) \equiv \phi_\infty/\sqrt{m_B}$
are concerned, there is a large spread amongst the results obtained by
different groups \uclafb
\ref\elcfb{C. Allton {\it et al.}, \NPB{349} (1991) 598
    \semi  S. Abada {\it et al.}, \NPB{376} (1992) 172.}
\ref\ukqcdfb{D. Richards {\it et al.} (UKQCD collaboration),
             in Ref. \amsterdam.}.
This discrepancy is likely to be due to inadequate isolation of the
lightest pseudoscalar meson state, and should shrink with the use of
optimised operators by all the groups. The uncertainty for the physical
$f_B$ and $f_D$ is, fortunately, much smaller. Some of the systematic
errors, such as uncertainties in the lattice scale $a$ and the axial
current renormalisation constant $Z_A$, can be cut down by expressing
the results as ratios of pseudoscalar decay constants. The quenched
lattice results then become \uclafb\elcfb\ukqcdfb:
$f_D/f_B$, $f_{B_s}/f_B$ and $f_{D_s}/f_D$ around $1.1$ with errors
less than $5\%$; while $f_B/f_\pi = 1.4(1)$.

A precise value for $f_B$ has important phenomenological implications.
Constraints from analyses of $K\bar{K}$ and $B\bar{B}$ mixing as well
as $b-$decays, together with the anticipation $m_t \ge 130$ GeV,
leave open two possibilities for the $CP-$violating phase $\delta$
in the CKM quark flavour mixing matrix
\ref\cpphase{See for instance: M. Lusignoli, L. Maiani, G. Martinelli
             and L. Reina, \NPB{369} (1992) 139.}.
The possibility of $\delta$ in the first(second) quadrant corresponds
to large(small) $CP-$violation in the decay $B \to K_S~J/\psi$.
The knowledge of $f_B$ can help distinguish between these two choices,
since $f_B/f_\pi = 1.5-2$ in the former case while $f_B/f_\pi \approx
1$ in the latter. The lattice results described above, provided that
the unknonwn systematic error due to the quenched approximation is
not too large, favour $\delta$ being in the first quadrant.

\newsec{$4-$Fermion Weak Interaction Matrix Elements} 

There are many situtations in the standard model where perturbation
theory in the electro-weak sector is adequate but non-perturbative
aspects of QCD substantially alter the amplitudes.
The usual machinery of operator product expansion, integration of heavy
gauge and matter fields, renormalisation group evolution and operator
mixing leaves one with low energy $4-$Fermion effective interactions
whose matrix elements have to be non-perturbatively evaluated
taking in to account all the QCD corrections.
Several of these amplitudes involve the (pseudo-)Goldstone bosons and so
chiral symmetry plays an important role in restricting the structure
of the amplitudes. Both Wilson and staggered fermion formulations have
been applied on the lattice to the study of the problem, and results
from the two approaches should agree for a consistent estimate.
Due to its residual chiral symmetry, the staggered fermion approach
has an inherent advantage; at present the best results for the weak
interaction matrix elements come from this approach.

\subsec{$K^0 - \bar{K^0}$ Mixing $B-$Parameter} 

This parameter characterises the magnitude of the $CP-$violation
parameter $\epsilon$, and is defined as
\eqn\bkprime{
B_K ~=~ { \vev{\bar{K^0}^\prime|
  \bar{s} \gamma_\mu (1-\gamma_5) d
  \bar{s} \gamma_\mu (1-\gamma_5) d |K^0} \over {8\over3}
  \vev{\bar{K^0}| \bar{s} \gamma_\mu\gamma_5 d |0}
  \vev{0| \bar{s} \gamma_\mu\gamma_5 d |K^0} } ~~.
}
It is the ratio of the true matrix element to its value in the Vacuum
Saturation Approximation. The lattice results show that the VSA does
not work for individual components of the matrix element. Indeed the
vector and axial parts separately are logarithmically divergent in the
chiral limit. It is only when all the pieces are combined together that
the net result is finite
\ref\bkone{G. Kilcup, S. Sharpe, R. Gupta and A. Patel,
           \PRL{64} (1990) 25.}.
Simulations using staggered fermions have been carried out on large enough
lattices and with sufficient statistical precision to expose underlying
systematic effects. The values obtained for the renormalisation group
invariant combination $B_K g^{-4/9}$ show a systematic decrease with
decreasing lattice spacing, and can be extrapolated to around $0.55-0.60$
for $a \to 0$
\ref\bkstag{S. Sharpe, R. Gupta and G. Kilcup, in Ref. \tsukuba, p.197.}.

Results for $B_K$ obtained using Wilson fermions have to undergo a
non-perturbative subtraction procedure to eliminate the chiral symmetry
violating lattice artifacts. Though the final value for $B_K$ again comes
out to be around $0.6$, the error on it is at least a factor of $5$
larger compared to the staggered fermion case \bkwil. Simulations using
an improved Wilson fermion action considerably eliminate the need for a
non-perturbative subtraction and hold a better promise for the future
\ref\bkimpwil{G. Martinelli, C. Sachrajda, G. Salina and A. Vladikas,
              \NPB{378} (1992) 591.}.

\subsec{$B-$parameters for Left-Right Operators} 

The parameter $\epsilon'$ characterises direct $CP-$violation in $K \to
2\pi$ decay amplitudes. In this case the matrix elements are those of
the imaginary part of the weak interaction Hamiltonian. The coefficient
functions in front of the effective operators depend on the top quark
mass and new physics, and a measurement can decide whether anything
beyond the standard model is necessary to understand $CP-$violation
\ref\cpepstop{G. Buchalla, A. Buras and M. Harlander,
              \NPB{337} (1990) 313.}.
In the currently allowed range of $m_t$ there are substantial
cancellations between various pieces contributing to $\epsilon'$.
Matrix elements of the QCD penguin operators are the main ingredients
to our understanding of the situation. Isospin violating corrections,
from unequal quark masses and electro-penguin operators, also make
significant contributions to $\epsilon'$. All the necessary matrix
elements correspond to $4-$Fermion operators with left-right chiral
structure. The lattice data show that they are not too far off from
from the VSA guess, i.e. $B_{LR} \approx 1$
\ref\bpenguin{S. Sharpe, in Ref. \tallahassee, p.429.}
\bkwil---a result that can be understood to some extent on the basis
of the rigorous inequalities of subsection 2.3.

\newsec{Sea Quark Content of the Hadrons} 

Most of the numerical results regarding the hadronic spectral parameters
show little difference between the quenched approximation and the full
theory (although the bare gauge couplings in the two cases are quite
different). Also the phenomenological quark models, which ignore the
sea quarks altogether, do a good job in fitting many of the experimental
results. All this suggests a welcome phenomenological simplification of
the theory (i.e. the dominant effect of the sea quarks is to generate
constituent quarks as valence quarks with renormalised properties),
but one would like to understand the dynamical reason behind it.

To find unambiguous signatures of the sea quarks (or failures of the
quenched approximation), one has to look for instances where the sea
quarks amount to more than mere renormalisation of the gauge coupling.
Good places to search are then the cases where the quark model
expectations are not a good guide.
Of particular interest are the strange quark bilinear matrix elements
of the proton $\vev{p| \bar s \Gamma s |p}$. Especially the scalar
density ($\Gamma=1$) and the axial current ($\Gamma=i\gamma_\mu\gamma_5$)
bilinears are connected to dynamically broken classical symmetries of QCD
and have a significant scope for mixing with the gluonic sector/sea quarks.

The sea quark matrix elements are technically more difficult to calculate
than the valence quark matrix elements. The main reason for this is that
the correlation of an insertion on the vacuum quark loop with the hadron
propagator, occuring via multiple gluon exchanges is statistically noisy.
It turns out that, instead of directly computing the $3-$point correlation
functions, such correlations are easier to extract by making the hadron
propagate through a background external field
(i.e. creating the sea quark configurations with an extra source term
$S_\Gamma = h_\Gamma \sum_x \bar{\psi}(x) \Gamma \psi(x)$
added to the standard fermion action), and then evaluating numerical
derivatives with respect to the external field strength $h_\Gamma$.

\subsec{The $\pi-N$ $\sigma-$term} 

The effect of dynamical quarks is clearly seen in the case of the
pion-nucleon sigma term (with $m=m_u=m_d$):
\eqn\sigmaterm{
\sigma_{\pi N} ~=~ m \vev{N| (\bar u u + \bar d d) |N}
               ~=~ m ({\partial \over \partial m_u}
                    + {\partial \over \partial m_d}) m_N ~~,
}
where the quark mass derivatives are to be evaluated at fixed gauge
coupling. The most recent analysis
\ref\sigmaexpt{J. Gasser, H. Leutwyler and M. Sainio,
               \PLB{253} (1991) 252.}
of experimental data (better data are definitely desirable) gives
$\sigma_{\pi N} \approx 45$ MeV. Within the first order flavour
$SU(3)$ breaking parametrisation, the valence quark component is
only $\sigma_{\pi N}^{\rm val} \equiv m(3F_S-D_S) \simeq 26$ MeV,
leaving ample room for the sea quark component
$\sigma_{\pi N}^{\rm sea} \equiv 2mS_S$.

Our Wilson fermion results
\ref\qcdwfb{R. Gupta, C. Baillie, R. Brickner, G. Kilcup, A. Patel
            and S. Sharpe, \PRD{44} (1991) 3272.}
for the ratio of the full matrix element to its valence part
show that the sea contribution is $1-2$ times the valence part.
When the staggered fermion results of the Columbia group
\ref\columbia{F. Brown {\it et al.}, \PRL{67} (1991) 1062.}
are analysed in a similar manner
\ref\seacontent{A. Patel, in Ref. \tsukuba, p.350.},
the contribution to $\sigma_{\pi N}$ from the sea and the valence
components is again found to be comparable. This feature is
qualitatively in agreement with the experimental data and gives an
indication of the importance of insertions on quark loops.
The overall magnitude of $\sigma_{\pi N}$ is systematically lower
than the experimental value in these simulations, however, an effect
likely to be due to not having explored small enough quark masses
and weak enough gauge couplings.

The analogous scalar density matrix elements of the $\rho$ and the
$\Delta$ show similar factors between the valence and the full value,
while the magnitude of the matrix elements is roughly proportional to
the number of valence quarks. This suggests a model for constituent
quarks in which the quarks are dressed strongly, and in a manner which
is independent of the state that they are in. Indeed, it can be reasoned
\ref\seasigb{A. Patel, \PLB{236} (1990) 102.}
that the large sea quark component we see is due to change in the overall
scale of the theory; the sea quarks influence the $\beta-$function
through vacuum polarisation. The really surprising feature of the lattice
results then is that even relatively heavy sea quarks (corresponding to
the pseudoscalar meson mass up to $1-1.2$ GeV) give a contribution to
scalar density matrix elements that is comparable to the valence component.

\subsec{Polarisation of Sea Quarks} 

The EMC result on polarised muon-proton scattering is difficult to
digest without a significant contribution from the sea quarks. Lattice
measurement of sea quark and gluon components of the structure function
$g_1$ can be attempted in two ways. The simpler approach is to replace
the insertion on the sea quark loops by an effective gluon operator.
The measurement of the non-forward matrix element of $\Tr(F{\tilde F})$
and extrapolation of the result to zero momentum transfer then gives
an estimate of the sea quark contribution to the axial current matrix
elements. Application of this method within the quenched approximation
at $\beta=5.7$ has given an upper bound $|S_A| < 0.08$
\ref\goneffdQ{J. Mandula, in Ref. \tsukuba, p.356.}.
A more sophisticated calculation using four flavours of dynamical
staggered fermions on $16^3 \times 24$ lattices at $\beta=5.35$ and
$ma = 0.01$ (corresponding to pseudoscalar mass of about $0.55$ GeV)
has produced the first tantalising result: $S_A = 0.05(1)$ per flavour
\ref\goneffdD{R. Altmeyer, M. G\"ockeler, R. Horsley, E. Laermann
              and G. Schierholz, in Ref. \amsterdam.}
\ref\goneffdtruth{Contrary to their claims, both Refs. \goneffdQ
                  \goneffdD\ extract only the sea quark contribution
                  to the singlet axial current matrix element and not
                  the full singlet axial current matrix element.}.

An alternative approach is to directly determine the axial current
coupling to the sea quarks in the proton. A hint for a possible signal
has been seen in a quenched approximation calculation
\ref\goneseaQ{J. Mandula and M. Ogilvie, PRINT-92-0341 (DOE),
              hep-lat 920809 (1992).}.
We have attempted a full theory simulation with two flavours of dynamical
Wilson fermions in a background singlet axial current field \seacontent\
\ref\goneseaD{R. Gupta, A. Patel and S. Sharpe, in progress.}.
In the first trial run, we used $8^4$ lattices at $\beta = 5.3$.
The quark mass parameters were $\kappa = 0.165, 0.166, 0.167$
with the corresponding background axial current field strengths
$2\kappa h_A = 0.005, 0.004, 0.003$. The pseudoscalar masses are
around $0.9, 0.8, 0.6$ GeV for these relatively heavy quarks.
With these lattice parameters, we have been unable to find any
numerical signal for $S_A$, and can express our results only as
$2\sigma$ upper bounds: $|S_A| < 0.05, 0.08, 0.11$ per flavour.

Assuming flavour $SU(3)$ symmetry, these numbers are a factor of $2$
to $3$ below what is required to confirm the EMC result on the lattice.
This feature sharply contrasts with the scalar density case, where sea
and valence quark components are comparable in magnitude at similar
quark masses. It is entirely possible that $S_A$ moves away from zero
at comparatively small quark masses only and consequently would be much
more difficult to measure.

Lattice results for non-singlet scalar density and axial current matrix
elements have turned out to be reasonable
\ref\nonsing{See for instance: Ref. \qcdwfb\ and references therein.}.
Combining all the observations for scalar density and axial current
couplings of the baryon octet with the heavy quark limits and
constituent quark model expectations we find:
(a) $F_S$, $mS_S$, $F_A$ and $D_A$ are finite in the heavy quark
limit and are smooth functions of the quark mass, while (b) $mD_S$
and $S_A$ vanish in the heavy quark limit as $m^{-2}$ and depart
significantly from zero only when the quark mass becomes of the order
of $\Lambda_{QCD}$. This suggests that flavour $SU(3)$ breaking effects
are probably small in the former case, though likely to be sizeable
in the latter case.

\bigskip
\line{\bf Acknowledgements \hfill}
\medskip

I thank Guido Martinelli for having pointed out to me the existence of
rigorous inequalities for the matrix elements of $4-$Fermion operators.
I am grateful to Stephen Sharpe for conveying to me many of the results
presented at LAT92.

\listrefs
\vfill\bye